\documentclass[aps,longbibliography,notitlepage,nofootinbib,11pt, floatfix,tightenlines%,superscriptaddress
]{revtex4-2}

\usepackage{amsmath}
\usepackage{amssymb}
\usepackage{bm}
\usepackage{graphicx}
\usepackage[format=plain,labelfont={bf,small},textfont=small,justification=raggedright,singlelinecheck=false]{caption}
\usepackage{url}
\usepackage{xcolor}
\usepackage{tensor} % mixed notation etc

\newcommand{\bx}{{\mathbf x}}

\newcommand{\ba}{{\mathbf a}}

\newcommand{\bn}{{\mathbf n}}

\newcommand{\bb}{{\mathbf b}}
\newcommand{\bB}{{\mathbf B}}

\begin{document}
\count\footins = 1000 %this partially fixes revtex bug where footnotes get cut off in onecolumn. however, it really screws up any color going between the two pages either in the footnote or the regular text

\title{Shear stresses in fluid and solid membranes with bending elasticity}

\author{S. Dharmavaram}
	\email{sd045@bucknell.edu}
	\affiliation{Department of Mathematics, Bucknell University, 1 Dent Drive, Lewisburg, PA 17837, U.S.A.}
\author{J. A. Hanna}
	\email{jhanna@unr.edu}
\affiliation{Department of Mechanical Engineering, University of Nevada, 1664  N.\  Virginia  St.\ (0312),  Reno,  NV  89557-0312,  U.S.A.}

\begin{abstract}
Comparison of a few simple models of fluid and solid membranes illustrates how shear stresses can arise from a bending energy through a coupling between curvature and surface stresses, a feature incidental to the fluid or solid nature of the material. In particular, it is shown how a fluid-like Helfrich bending energy contributes shear stresses, while a related solid-like energy, the correct continuum limit of a Seung-Nelson discrete bending term, produces a stress tensor with purely isotropic tangential part. 
A distinction is noted between the resistance to tangential flows on the surface and the ability to support tangential stresses. The tangential projection of the divergence of the stress, or the pseudomomentum balance, is viewed in the light of some statements and observations in the literature. 
Different types of constraints, invariance properties, and shape equations are briefly discussed.   
The state of a patch of unidirectionally-bent material is connected with prior analyses of the pulling of membrane tethers. 
\end{abstract}

% \pacs{}

\date{\today}

\maketitle

%alternative constitutive choices

\section{Introduction}\label{intro}

The relation between the elastic energy and the resulting stresses in thin bodies is not simple. Constitutive choices at the level of the energy can have non-intuitive consequences in the balance equations. 
The shear stresses arising in fluid membrane models, such as those commonly applied to lipid bilayers, present one example.  
The origin of these shear stresses, which some in the soft matter community find conceptually troublesome, lies in certain aspects of the bending energy that are only incidentally related to its fluid nature.  
%doesn't come from geometry, but from constitutive choice. 

In this paper, we consider bending energies extracted from two related membrane models commonly used in soft matter research, the Helfrich energy \cite{Helfrich73} based on squared mean curvature, and the continuum limit of the discrete Seung-Nelson (SN) energy \cite{SeungNelson88} based on differences between normals on a triangular mesh.  
One is a fluid model that supports shear stresses, and the other is a solid model that does not--- to be precise, the bending portion of the model does not support them, but it is usually accompanied by an additional stretching energy that does. 
For further illustration, we pair these bending energies with either a fluid-like constraint on area or a solid-like constraint on the metric. 
%We observe that 
Physicists often incorrectly identify the SN bending energy with Helfrich, an approximation that applies in the limit of small strain, %and small curvature,
 only enforceable for sufficiently stiff or fully metrically-constrained solid membranes. 
%helfrich vs SN: difference goes like Y*strain plus B*curvature.  The latter can be important esp for fluid membrane, where it is the only contribution to shear resistance. 
As a result, the continuum form of this model has not been thoroughly examined.\footnote{The name Seung-Nelson follows common convention in the soft matter field, but perhaps a more historically accurate name would be the Kantor-Nelson energy \cite{KantorNelson87prl}.  Kantor and Nelson make the reasonable identification of their continuum limit with a small-strain, small-slope approximate plate energy at lowest order in displacement derivatives, where the distinction between present and referential gradients and metrics is small.
 %which is fine as at that order (low strain and curvature) the distinction in gradients is ignorable.  
Seung and Nelson incorrectly infer a more general exact correspondence between the difference in normals and the (present) gradient of the normal field, and thereby with Helfrich. 
%\comb{``in the continuum limit the difference [ na-nb ] should become the gradient of the normal vector field ... [ integral expr with $g^{ij}$dindjn ] ''}
Seen within the context of their broad and insightful paper, the error is of minor and incidental significance, but the paper has been very influential; the downstream effects of this misidentification have at times been important, for example whenever the coupling between curvature and tension is of interest. An exhaustive list of soft matter papers where this error has propagated is impractical here. It makes a notable appearance in an early high-profile publication \cite{KrollGompper92} (despite its title, this paper does not model a fluid membrane, but a bead-spring SN membrane), and again in a review by the same authors \cite{GompperKroll97}. 
%, which also makes the same shear stress error noted below
Many other works repeat the error, while many others do not explicitly make the error, but do not clearly address the approximation.}

%Note is inspired by our repeated encounters with statements to the effect that 
That a fluid model can generate shear stresses has been a common point of confusion in the literature.  
While it is well known that the Helfrich energy produces an anisotropic tangential contribution to the membrane stress tensor, as derived for example in \cite{Jenkins77siam, Steigmann99arma, RossoVirga99, CapovillaGuven02, LomholtMiao06, FengKlug06}, 
many users of this energy are uncomfortable identifying these terms as shear stresses, because of a misperception that such quantities cannot be supported by a fluid membrane (examples are provided below in Section \ref{fluidity}).  
There are two main points we wish to make.   
The first is that the quantities appearing are indeed shear \emph{stresses} that can resist forces tangential to the membrane, but they arise from the resistance to bending, that is, deformations of the shape of the membrane, and not to shear \emph{deformations within} the membrane that tangentially displace material. %confusion between deformations within, and deformations of, the surface. 
In a fluid film with bending elasticity, only the former deformations contribute to the energy. 
% in which material moves tangentially
This identification of shear stresses is rarely stated explicitly (exceptions are provided below in Section \ref{fluidity}). 
The second point is that shear or any other tangential stresses are not some inevitable consequence of ``geometry'', but need not arise at all from a bending energy.  
The appearance and sign of these terms are due to the dependence of the energy on the stretch, the tensor encoding the ratio of present to referential lengths, which dependence governs the coupling between bending and tangential stresses \cite{IrschikGerstmayr09, OshriDiamant17, WoodHanna19}. 
In particular, the fluid Helfrich bending energy displays this coupling, while the solid SN bending energy has a more primitive constitutive response that does not give rise to shear or any other tangential stresses (this is shown in Section \ref{energies}; what few new calculations are needed are presented in Appendix \ref{SNstress}, and a comparison for a unidirectionally-bent patch of material, as in certain models of pulled membrane tethers, is presented in Section \ref{rudimentary}).  % (Appendix \ref{SNstress}). 
The stretch is a quantity relevant in elasticity rather than fluid mechanics, and the insight just mentioned arises unexpectedly from considerations natural to the elasticity of thin solid bodies that may seem foreign to the study of fluid membranes. 
Section \ref{eshelby} explores the identification of projections of the momentum balance equations with the 
%Eshelby tensor and
 pseudomomentum balance, noting in passing how certain rearrangements of terms, made possible by geometric relations, mask some of the structure and relationships present. 

We have intended to write these observations with the primary perspective, notation, and intended audience of physics, with occasional mention of connections to continuum mechanics when warranted.

\section{Fluidity and its malcontents}\label{fluidity} 

The energy of an incompressible fluid surface is invariant under area-preserving diffeomorphisms of some chosen reference configuration, or equivalently under the unimodular group of transformations of its tangent space  \cite{Jenkins77siam, Steigmann99arma}.    Physically, this corresponds to deformations consisting of area-preserving, shape-preserving (tangential) rearrangements of material on the surface.  This fluidity is distinct from the ability to reparameterize (relabel) material elements in a homogeneous medium. The latter property is not unique to fluids, and gives rise to the conservation of pseudomomentum discussed in Section \ref{eshelby}. 
% relabeling symmetry %\cite{padhyemorrison1996}, often called reparameterization invariance, 
As shear deformations within a fluid surface do not cost energy, they produce no static forces, and will be resisted only by viscous forces if present. 
By contrast, rearrangements % (but not reparameterizations)?? 
of a solid membrane are tightly constrained to preserve a two-dimensional metric, not just an area. %killing field

The misinterpretations in the literature are a conflation of the cost of shearing motions and the ability of the membrane to support static shear stresses, as are present in an anisotropic tangential part of the stress tensor. 
It is correct to say that tangential displacements of fluid elements within the membrane do not cost energy, and thus offer no static resistance \cite{Jenkins77jmb, CapovillaGuven04, ArroyoDeSimone09}. 
%\comb{Jenkins77jmb "Shear deformations in the surface of the membrane are resisted only by viscous forces" CapovillaGuven04: "the vesicle acts like a two-dimensional fluid as there is no cost in energy associated with tangential displacements of the lipid constituents." ArroyoDeSimone09: "invariance under the action of tangential velocity fields whose normal component to the surface boundary vanish since these do not change the surface geometry." %seem to also identify shears at boundary, but not very clear.  also seem to conflate reparameterization, etc.}
Indeed this lack of penalty for flows on the surface leads to numerical challenges, particularly with the use of finite element methods. 
A singular energy Hessian with zero-energy modes gives rise to severe mesh distortions \cite{FengKlug06}, which have been addressed by adding in-plane viscosity \cite{MaKlug08} (see also \cite{zarda1977elastic}) or other techniques that do not change the constitutive law 
\cite{FengKlug06, zhao2017direct, sahu2020arbitrary, Dharmavaram21};  
% restricting the representation/Monge \cite{zhao2017direct} and others? 
note that \cite{FengKlug06, MaKlug08, Dharmavaram21} conflate this fluidity with reparameterization invariance. 
However, this does \emph{not} mean that the surface cannot support tangential shear forces, an incorrect remark that appears repeatedly in the literature. 
While no stress is associated with moving fluid tangentially within the membrane, the membrane can still sustain a tangential shear stress. The surface can change shape, and for \emph{some} choices of bending energy this will give rise to tangential stresses which are, in general, anisotropic. Conversely, such a bent membrane will have tangential stresses that must be balanced by other things, such as applied forces. 
%\com{give/describe example of applied shear force and bending response? unidirectional bending with shear forces applied at some angle to the bending}
This dependence of tangential stresses on the choice of bending energy will be illustrated in Section \ref{energies}. 

Some researchers have recognized, at times reluctantly, the existence of shear (anisotropic) stresses. 
Powers and co-workers \cite{Powers02} begin by considering only the normal part of the momentum balance.  % the shape equation (normal forces) and do not derive tangential stress balance. 
Their assumption that a fluid membrane must have isotropic tension leads to an ``apparent paradox'' in the modeling of a membrane tether.  Upon further comparison with force and moment balances for axisymmetric, nearly flat plates, they conclude that the bending moments contribute an additional ``tension'' (quotes theirs), and suggest a possible physical mechanism for its origins. 
%\comb{``...an apparent paradox arises'' because they assume the tension in the membrane ''is isotropic, since the membrane is fluid.'' but by thinking about axisymmetric balances for plates they conclude in appendix that "Part of the ‘‘tension’’ in the membrane comes from the bending moments."  Note that their shell balance analysis is based on EvansYeung94 who seem to invoke simply "a small lateral projection of transverse shear due to curvature" and assume rest of tension is isotropic. } 
Others derive tangential stresses from variation of the energy, in which anisotropic terms clearly appear. 
Tu and Ou-Yang \cite{TuOuYang08} state that ``we seem to arrive at a paradox... fluid membranes cannot withstand in-plane shear strain, however... % Eqs. (60) and (61) reveals 
shear stress still exhibits in non-spherical vesicles.''  
Guven and V{\'{a}zquez-Montejo \cite{GuvenVMinbook} state that ``There will generally be a geometrical in-plane shear [force]... There is, of course, no inconsistency with the fluid character of the membrane.'' %which seems to be correct but does not warrant an ``of course''. 
Sauer and co-workers \cite{Sauer17} state that ``Interestingly, the bending part of the Helfrich model can contribute an in-plane shear stiffness''.  
Other, and some of the same, authors make incorrect statements regarding stresses, namely that the membrane offers no resistance to shear forces, or that what appear to be shear forces cannot actually be, because the membrane is fluid \cite{GompperKroll97, FengKlug06, MaKlug08, Maleki13, GuvenVazquez-Montejo13, Deserno15}.  
%\comb{GompperKroll97: "... the fluid structure, which does not allow a preferred coordinate system, and therefore cannot support shear stress". FengKlug06: "bilayer membranes are only able to support external forces applied in the direction normal to the surface, and not tangentially applied forces." they are getting confused with the shearing flows they see. similarly first sentence of abstract of MaKlug08: ``As two-dimensional fluid shells, lipid bilayer membranes resist bending and stretching but are unable to sustain shear stresses.'', `` there is no shear force because of the fluid property of membranes.'', Maleki13: "...a lipid bilayer in the liquid phase does not have the ability to resist in-plane shear forces. This is because the lipid molecules may move freely within a lipid bilayer." GuvenVazquez-Montejo13: footnote "There is, of course, no shear stress supported by a fluid membrane. A nondiagonal tensor will occur when the tensor is decomposed with respect to an inappropriate frame." review Deserno15 footnote: ``Given that the off-diagonal component of the curvature tensor is involved, it is tempting to interpret this force as a shear, but this view is misleading: The surface is fluid and cannot support shear.'' }
Yet other, and some of the same, authors make ambiguous statements whose correctness depends on interpretation \cite{Deseri08, Capovilla17, Sauer17}. 
%\comb{Deseri08 `` the membrane behaves like an in-plane-fluid, which roughly means that lipid molecules can move freely on the membrane surface. From the mechanical point of view, in-plane-fluidity means that the membrane does not offer resistance to shears contained in planes orthogonal to [the referential normal].''  if deformations ok if stresses no. 
% define no shears in 3d despite saying 'in-plane fluidity'. don't revisit after reduction to 2d.  no '2d fluid' ?early in Sauer17 paper it confusingly says "Ideal liquids lack shear stiffness. Under quasi-static conditions, liquid membranes and shells therefore do not provide any resistance to in-plane shear deformations and thus need to be stabilized." Sauer's contribution to Steigmann's edited book says same stuff.  also Capovilla17 reviewish "fluidity of the membrane, or negligible shear, ..."} 
This representative sample of the works of prominent researchers in the field, including review articles and contributions to the 2016 CISM Advanced Summer School on “The Role of Mechanics in the Study of Lipid Bilayers”, should indicate the prevalence of the issue. 
%These are only a representative, not comprehensive, sample of the works of prominent researchers in the field, which % but its inclusion of review articles and invited contributions to the 2016 CISM Advanced Summer School on “The Role of Mechanics in the Study of Lipid Bilayers”
% should indicate the prevalence of the issue. 

%\comb{origins of shear stresses misinterpreted: Powers02 invoke cartoon argument "Axial force due to the greater circumference of the outer leaf."  ... "The axial force at zero pressure jump is a manifestation of the liquid properties of lipid bilayer membranes." (no, it's a manifestation of the choice of mean curvature squared). GuvenVM in book comments: "The structure captured in Eqs. (51) and (52) is independent of the specific form of H . As promised, the stress is completely determined by the geometry. This is quite unlike the familiar situation in continuum mechanics where in-plane static shear— which is not supported by a two-dimensional incompressible fluid—generates stress." but this is a bit misleading.  stress does depend on specific form of H even if the structure description is universal. }

\section{notation, definitions, and tricks}\label{notation}

%We denote material coordinates as $\eta^\alpha$, $\alpha \in \{1,2\}$ and 

We denote %(noncovariant material) 
 derivatives with respect to material coordinates as $d_\alpha$, $\alpha \in \{1,2\}$. 
These coordinates parameterize a surface %$\bx(\eta^\alpha)$
$\bx$ in $\mathbb{E}^3$ with normal $\bn$, tangents $\ba_\alpha = d_\alpha\bx$, reciprocal tangents defined through the relations $\ba^\alpha\cdot\ba_\beta = \delta^\alpha_\beta$, metric components $a_{\alpha\beta} = \ba_\alpha\cdot\ba_\beta$ and $a^{\alpha\beta} = \ba^\alpha\cdot\ba^\beta$, area form $\textrm{d}a$, covariant derivative $\nabla_\alpha()$ constructed with the metric, and curvature components $b_{\alpha\beta} = b_{\beta\alpha} = d_\beta \ba_\alpha\cdot\bn = -\ba_\alpha \cdot d_\beta\bn$. 
Indices on these ``present'' quantities are raised and lowered with the metric. 
We denote the surface gradient and Laplacian as $\nabla() = \nabla_\alpha()\ba^\alpha$ and $\nabla^2 = \nabla^\alpha\nabla_\alpha$, and the mean and Gau\ss ian curvature invariants as $H = \tfrac{1}{2}b_\alpha^\alpha$ and $K = \tfrac{1}{2}\left(b_\alpha^\alpha b_\beta^\beta - b^\alpha_\beta b_\alpha^\beta\right)$. 

Similar quantities can be defined for a reference surface parameterized by the same coordinates.  Of these we will only require
the area form $\textrm{d}A$, metric components $A_{\alpha\beta}$ and $A^{\alpha\beta}$, and covariant derivative $\bar\nabla_\alpha$ constructed with the reference metric. 
Indices on these ``referential'' quantities are raised and lowered with the reference metric. 
The area forms are related by $\textrm{d}a = J\textrm{d}A$, where the Jacobian determinant $J$ is the ratio of present to referential metric determinants.  % the examples in this paper, $J=1$. 
 
Derivations will make repeated use of the following facts. Derivatives $d_\alpha$, $\nabla_\alpha$, and $\bar\nabla_\alpha$ act identically on objects without free indices.  Two derivatives of any type acting on an object without free indices can have their indices (not type) permuted, as the connections are torsion-free (Christoffel symbols are symmetric). %, for example $\nabla_\alpha d_\beta \delta\bx = \nabla_\beta d_\alpha \delta\bx$. 
The surface $\bx$ obeys the Weingarten relations $\nabla_\beta\ba_\alpha = b_{\alpha\beta}\bn$ and $d_\alpha\bn = -b_{\alpha\beta}\ba^\beta$ and the Codazzi relations $\nabla_\gamma b_{\alpha\beta} = \nabla_\alpha b_{\gamma\beta}$ or, equivalently, $\nabla_\gamma (d_\alpha\bn\cdot\ba_\beta)  = \nabla_\alpha (d_\gamma\bn\cdot\ba_\beta)$. The Piola identities in component form are $\bar\nabla_\alpha ()^\alpha = J\nabla_\alpha(J^{-1}()^\alpha)$ and $\nabla_\alpha ()^\alpha = J^{-1}\bar\nabla_\alpha(J()^\alpha)$, and take a particularly trivial form when $J=1$, as will be true for all of our examples.

\section{energies and stresses}\label{energies}

Let us compare five expressions representing three different constrained energies, constructed using three types of bending content and two types of constraint. 
We are primarily interested in two of these expressions; the others are included for further illustration and deflection of potential questions.  
For simplicity of exposition, we consider membranes with vanishing rest curvature. Either constraint 
will prevent local area change, so that $J=1$ and thus $\mathrm{d}a = \mathrm{d}A$. 

The first two expressions we will refer to as Helfrich and referential-Helfrich, with the local area constraint $J=1$: % (as written it could be just a surface tension, but we can presume a constraint $J=1$):
\begin{align}
	\mathcal{E}_\text{\tiny (H)} &= \int \!\mathrm{d}a \left( \tfrac{1}{4}\nabla^\alpha \bn \cdot \nabla_\alpha \bn + \sigma_\text{\tiny (H)} \right) \nonumber \, , \\
	&= \int \!\mathrm{d}a \left( H^2 - \tfrac{1}{2}K + \sigma_\text{\tiny (H)} \right) \nonumber \, , \\
	&= \int \!\mathrm{d}A\, J \left( H^2 - \tfrac{1}{2}K + \sigma_\text{\tiny (H)} \right) \label{helfrich} \, , \\
	\mathcal{E}_\text{\tiny (rHa)} &= \int \!\mathrm{d}A \left( H^2 - \tfrac{1}{2}K  + J \sigma_\text{\tiny (rHa)} \right) \, . \label{rhelfricha}
\end{align}
Subscripts in parentheses are not indices. 
%The subscripted parentheses are a reminder that their contents are not indices
%The parentheses are a reminder that the enclosed subscripts are not indices.  
%where $H^2 - \tfrac{1}{2}K =  \frac{1}{4}\nabla^\gamma \bn \cdot \nabla_\gamma \bn $ %noting that $\nabla\bn = d_\alpha\bn\, \ba^\alpha$ is symmetric. 
Because of the constraints, these expressions represent the same energy, so the resulting shapes and stresses will be the same, but this requires %specification of 
different scalar multipliers $\sigma_\text{\tiny (H)}$ and $\sigma_\text{\tiny (rHa)}$. 
Outside of the soft matter community, the bending part of \eqref{helfrich} is often associated with the name of Willmore \cite{PinkallSterling87}. 
The full integral of this bending content--- not its local density--- is conformally invariant up to a boundary term \cite{White73}, although this is not of much relevance in the presence of the area constraint. 
Besides inclusion of a referential curvature, Helfrich can also have different moduli for $H$ and $K$. The choice made here allows for easy comparison with SN and other elastic models, and is irrelevant to the points made in this paper. 
It is also common to consider a global constraint on area.  
In computations, it is more convenient to impose either global area constraints or a penalty term to avoid mixed finite element formulations, which are often sensitive to numerical instabilities.  We will not delve into these options here. 

The model (\ref{helfrich}-\ref{rhelfricha}) is acceptable as a fluid model.  
As shown in \cite{Steigmann99arma}, in the absence of referential curvature, the invariance requirements of fluidity are that the energy density be a function of the geometric curvature scalars $H$ and $K$ and the Jacobian determinant $J$--- the only dependence on the referential metric is through the change in area.

The third expression is the continuum limit of the bending part of the SN energy, combined with the local area constraint $J=1$:
 %for ``everse'' of the tensor which would give the same energy 
\begin{align}
	\mathcal{E}_\text{\tiny (SNa)} &= \int \!\mathrm{d}A \left(  \tfrac{1}{4}\bar\nabla^\alpha \bn \cdot \bar\nabla_\alpha \bn  + J \sigma_\text{\tiny (SNa)} \right) \, . \label{SNa}
\end{align}
One should contrast
$\nabla^\alpha \bn \cdot \nabla_\alpha \bn = \nabla^\alpha \bn \cdot d_\alpha \bn %= 4H^2 - 2K
 = a^{\alpha\beta}d_\alpha\bn \cdot d_\beta\bn = a^{\alpha\beta}a^{\gamma\eta}b_{\alpha\eta} b_{\gamma\beta} = b^\alpha_\gamma b^\gamma_\alpha$ 
from Helfrich with $\bar\nabla^\alpha \bn \cdot \bar\nabla_\alpha \bn = \bar\nabla^\alpha \bn \cdot d_\alpha \bn = A^{\alpha\beta}d_\alpha\bn \cdot d_\beta\bn = A^{\alpha\beta}b_\alpha^\gamma b_{\gamma\beta} = A^{\alpha\beta}a^{\gamma\eta}b_{\alpha\eta} b_{\gamma\beta} = A^{\alpha\beta}a_{\gamma\eta}b_\alpha^\eta b^\gamma_\beta$ from SN (being careful to remember that indices contracted between $A^{\alpha\beta}$ and the other objects cannot be simultaneously raised and lowered).   
A physicists' interpretation, namely the use of present or referential metric to raise indices on derivatives of the normal, can also be expressed in continuum mechanical language. 
As the covariant components of the metric go like the square of stretch, comparison of the last terms shows that the SN energy goes like the square of the product of stretch and curvature,\footnote{Continuum mechanicians might express the SN bending energy as the trace of either the present tensor $\bb \cdot \bB \cdot \bb$ or, most likely, the referential tensor ${\bf{F}}^\top \cdot \bb^2 \cdot {\bf{F}}$, where $\bb$ is curvature, $\bB$ is the left Cauchy-Green deformation, and ${\bf{F}}$ is the deformation gradient, standard quantities that we do not define here.   %, (although they may not be fond of the dot product notation)
Similar re-expressions of quantities appearing in the Helfrich energy can be found in \cite{Steigmann99arma}. 
%\comb{Steigmann99arma talks about ${\bf{F}}^\top \cdot \bb \cdot {\bf{F}}$ -- later showing that its trace is $J^2H$-- and then derives fluid-appropriate quantities using that and C... in the end everything has to do with H, K, J}  
} rather than the square of curvature. 
The stretch dependence of the energy governs the coupling between bending and tangential stresses \cite{IrschikGerstmayr09, OshriDiamant17, WoodHanna19}. 

We suggest naming this continuum bending content the Atluri bending energy, as it forms the relevant part of the energy appearing in \cite{atluri1984alternate}. 
It is also the trace of the tensorial bending measure recently proposed by Virga \cite{Virga24}, and  
%It can also be obtained from tensorial measures, either that proposed by 
forms part of the energy constructed in \cite{vitral2023dilation, vitral2023energies}, the other part not being easily expressible in terms of metrics and curvatures. % \com{can it be derived just using the lurie tensor?} 
% many in soft matter work with SN which is a discretized version of the Atluri bending energy plus some in-plane stretching parts.   
Unlike curvature, the local bending content of \eqref{SNa} is dilation (scale) invariant \cite{Ghiba21, vitral2023dilation, vitral2023energies}; while this seems to be not of much relevance in the presence of the area constraint, it is related to the absence of coupling with in-plane stresses \cite{vitral2023dilation, vitral2023energies}. 

The model \eqref{SNa} cannot represent a fluid. The bending content depends on the reference metric, not just its area. 
This is true whether or not it is further combined, as it usually is, with other solid-appropriate terms penalizing in-plane deformations. 

The expressions/energies \eqref{helfrich} and \eqref{SNa} are of primary interest to us.  It is also illustrative to consider two more expressions representing one more solid energy, obtained by combining the referential-Helfrich or Atluri/SN bending content with the local metric constraints 
 $a_{\alpha\beta} = A_{\alpha\beta}$ appropriate to an inextensible solid surface \cite{GuvenMuller08}: 
\begin{align}
	\mathcal{E}_\text{\tiny (rHm)}  &= \int \!\mathrm{d}A \left( H^2 - \tfrac{1}{2}K + \tfrac{1}{2}\sigma_\text{\tiny (rHm)}^{\alpha\beta} \ba_\alpha \cdot \ba_\beta \right) \, , \label{rhelfrichm}\\
	\mathcal{E}_\text{\tiny (SNm)}  &= \int \!\mathrm{d}A \left( \tfrac{1}{4}\bar\nabla^\alpha \bn \cdot \bar\nabla_\alpha \bn  + \tfrac{1}{2}\sigma_\text{\tiny (SNm)}^{\alpha\beta}  \ba_\alpha \cdot \ba_\beta \right) \, . \label{SNm}
\end{align}
Because of the constraints, these expressions represent the same energy, so the resulting shapes and stresses will be the same, but this requires %specification of 
different tensor multipliers $\sigma_\text{\tiny (rHm)}^{\alpha\beta}$ and $\sigma_\text{\tiny (SNm)}^{\alpha\beta}$. 
Similarly, they are equivalent to the Guven-M{\"{u}}ller functional \cite{GuvenMuller08}.  In the context of an environment where one might find a lipid membrane, they might represent an unstretchable capsule. %\com{note the capsule/vesicle literature often says inextensible to mean area-inextensible, not fully inextensible}.  

The energy \eqref{SNa} is not the same as either of the identical pairs (\ref{helfrich}-\ref{rhelfricha}) and (\ref{rhelfrichm}-\ref{SNm}).  It is a rather strange construct, a combination of a solid-like bending energy and a fluid-like area constraint.  
It is used here merely to illustrate the absence of shear stresses from the bending component of the energy.  A solid membrane model would normally include not just bending but stretching terms, which would provide shear resistance, and would resemble the inextensibility constraints in model (\ref{rhelfrichm}-\ref{SNm}) in the limit of vanishing strain. 

Our focus is on the stress ${\mathbf f}^\alpha$ obtained from variation of the energy with respect to the position $\bx$ as, for example, in the referentially-defined 
\begin{align}
	\delta\mathcal{E} = \int \!\mathrm{d}A \left[ \bar\nabla_\alpha \left( {\mathbf f}^\alpha\cdot\delta\bx + {\bm{\mu}}\cdot\bar\nabla^\alpha\delta\bx \right) - \bar\nabla_\alpha{\mathbf f}^\alpha\cdot\delta\bx \right] \, ,	\label{variationgeneral}
\end{align}
following a trivial adaptation of the widely adopted formalism of Capovilla and Guven \cite{CapovillaGuven02}. 
%\cite{LomholtMiao06, Powers10, Deserno15}
In all of our examples, $J=1$ and therefore, by the Piola relation, the divergences are identical in component form, $\bar\nabla_\alpha{\mathbf f}^\alpha = \nabla_\alpha{\mathbf f}^\alpha$.  Many of the relevant calculations have been performed elsewhere \cite{Jenkins77siam, Steigmann99arma, RossoVirga99, CapovillaGuven02, LomholtMiao06, FengKlug06}.   Appendix \ref{SNstress} provides the remaining variation of the Atluri/SN bending energy found in \eqref{SNa} and \eqref{SNm}.  
The stresses corresponding to the five expressions (\ref{helfrich}-\ref{SNm}) are: 
\begin{align}
	\mathrm{with}\; J=1\,, \;\;\, \;\;\, {\mathbf f}^\alpha_\text{\tiny (H)} &= H\nabla^\alpha\bn - \nabla^\alpha H \bn + \left(\sigma_\text{\tiny (H)} + H^2\right)\ba^\alpha \, , \label{fhelfrich}\\
	\mathrm{with}\; J=1\,, \;\;\, \, {\mathbf f}^\alpha_\text{\tiny (rHa)} &= H\nabla^\alpha\bn - \nabla^\alpha H \bn + \left(\sigma_\text{\tiny (rHa)} + \tfrac{1}{2}K\right) \ba^\alpha \, , \label{frhelfricha} \\
	\mathrm{with}\; J=1\,, \;\;\,  {\mathbf f}^\alpha_\text{\tiny (SNa)} &= \quad\quad  - \tfrac{1}{2}\nabla_\gamma\left( A^{\beta\gamma}b_\beta^\alpha \right)\bn + \sigma_\text{\tiny (SNa)}\ba^\alpha \, , \label{fSNa} \\
	\mathrm{with}\; a_{\alpha\beta}=A_{\alpha\beta} \,, \;\;\, {\mathbf f}^\alpha_\text{\tiny (rHm)} &= H\nabla^\alpha\bn - \nabla^\alpha H \bn + \left(\sigma_\text{\tiny (rHm)}^{\alpha\beta} +\tfrac{1}{2}Ka^{\alpha\beta}\right) \ba_\beta \, , \label{frhelfrichm}\\
	\mathrm{with}\; a_{\alpha\beta}=A_{\alpha\beta} \,, \;\;\, {\mathbf f}^\alpha_\text{\tiny (SNm)} &= \quad\quad\quad\, - \,\nabla^\alpha H \bn + \sigma_\text{\tiny (SNm)}^{\alpha\beta} \ba_\beta \, , \label{fSNm}
\end{align} 
where we have used the isometric constraint (and Codazzi) to simplify the expressions in \eqref{fSNm}. 
As the first two area-constrained energies (\ref{helfrich}-\ref{rhelfricha}) are the same, we identify %(specify the forms?)
 $\sigma_\text{\tiny (H)} + H^2 - \tfrac{1}{2}K = \sigma_\text{\tiny (rHa)}$ in (\ref{fhelfrich}-\ref{frhelfricha}) in order that ${\mathbf f}^\alpha_\text{\tiny (H)}  = {\mathbf f}^\alpha_\text{\tiny (rHa)}$.  As the last two metric-constrained energies (\ref{rhelfrichm}-\ref{SNm}) are the same, we identify (using Weingarten)  $\sigma_\text{\tiny (rHm)}^{\alpha\beta} + \tfrac{1}{2}Ka^{\alpha\beta} - Hb^{\alpha\beta} = \sigma_\text{\tiny (SNm)}^{\alpha\beta}$ in (\ref{frhelfrichm}-\ref{fSNm}) in order that ${\mathbf f}^\alpha_\text{\tiny (rHm)}  = {\mathbf f}^\alpha_\text{\tiny (SNm)}$, noting further that the metric constraint implies $K=0$.  The $K$ can also be hidden in area-constrained models such as \eqref{rhelfricha}/\eqref{frhelfricha} where $\mathrm{d}A = \mathrm{d}a$ by using the fact that the variation of $\int \!\mathrm{d}a\, K$ is a pure surface divergence, % \com{if need cite? can use Capovilla02-4, Deserno15, Dias11, Hanna19}, 
redefining accordingly the relevant piece of the energy.

The stress \eqref{fSNa} arising from the
%tangent normal $\bm{\nu} = \nu_\alpha\ba^\alpha$
 area-constrained Atluri/SN energy %${\mathbf f}^\alpha_\text{\tiny fSN}$ 
is tangentially isotropic.  Aside from the normal term, its contraction with a tangent normal at the boundary (for example, $\nu_\alpha{\mathbf f}^\alpha$ where $\nu_\alpha$ is a component of the present tangent normal $\nu_\alpha\ba^\alpha$) provides a purely tangent normal (pressure-like) term and thus offers no resistance, whether through in-plane or bending deformations, to application of tangent tangent (shear) forces.  
In contrast to the tensor multipliers in the two metric-constrained models (\ref{frhelfrichm}-\ref{fSNm}), the scalar multiplier in the three area-constrained models (\ref{fhelfrich}-\ref{fSNa}) cannot absorb a tangentially anisotropic $- Hb^{\alpha\beta}$ term that appears in the Helfrich-type forces \eqref{fhelfrich}, \eqref{frhelfricha}, and \eqref{frhelfrichm} and provides a tangent tangent contribution at the boundaries.   
 This is the difference between constraining the area \eqref{fSNa} and constraining the entire metric \eqref{fSNm}. These two models differ, as only the latter can provide constraint shear stresses. 
 Section \ref{rudimentary} compares \eqref{fhelfrich} and \eqref{fSNa} for a rudimentary example, making use of further results from the following Section \ref{eshelby}. 
 
The simple appearance in component form of the Atluri/SN stress \eqref{fSNa} is deceptive.  This is a hybrid quantity, of surprising complexity given the ease of computation of the discrete SN energy. 
As written, one applies the present covariant derivative (Christoffel symbols constructed with the present metric) to a quantity that ``contracts'' (yet the contracted indices cannot be simultaneously raised and lowered) the contravariant components of the referential metric (indices raised with the referential metric) with the mixed components of the curvature (index raised with the present metric). 
Some of this complexity is hinted at in Appendix \ref{SNstress}, where appear expressions involving a referential covariant derivative acting on a present tangent vector.\footnote{Wrapping these quantities in index-free tensorial notation requires objects such as the two-point rotation or shifter tensors \cite{VitralHanna22}. Physicists might be comfortable relating the latter to dot products between referential and present tangent vectors, a process some continuum mechanicians would reject as a mathematical abomination.    } 
In fact, the Atluri/SN energy is related to other quantities with similar stretch and curvature dependence that arise in elastic theories but are not easily expressible in terms of metrics and curvatures \cite{vitral2023dilation, vitral2023energies}.

\section{projections and material stresses}\label{eshelby}

It is common to examine the normal and tangential projections not just of the stress ${\mathbf f}^\alpha$, but of its surface divergence $\nabla_\alpha {\mathbf f}^\alpha$ that contributes to the balance of momentum.  The normal projection of the divergence provides the ``shape equation'' of the membrane, although this name is somewhat misleading given that shape changes can couple to in-plane shear forces, as in the Helfrich model.    
The tangential projection can also be written as a surface divergence after suitable rearrangement \cite{GuvenMuller08}. It often goes unrecognized or unremarked that this divergence contributes to the balance of pseudomomentum, and its vanishing reflects the material symmetry of a homogeneous medium \cite{SinghHanna21}. 
%\cite{SinghHanna21,*SinghHanna21correction}

The divergence of the Helfrich stress \eqref{fhelfrich} yields
\begin{align}
	-\nabla_\alpha {\mathbf f}^\alpha_\text{\tiny (H)}  = \left[ \nabla^2 H + 2H(H^2-K - \sigma_\text{\tiny (H)}) \right] \bn  - \nabla\sigma_\text{\tiny (H)} \, , \label{divH}
\end{align}
making use of the identity 
$-H\nabla^2\bn\cdot\ba_\beta = -H\nabla_\alpha\left(\nabla^\alpha\bn\cdot\ba_\beta\right) = \nabla_\beta H^2$, a result of the (Weingarten and) Codazzi relations. 
Note the related identities $\nabla_\alpha Hb^{\alpha}_{\beta} = \nabla_\alpha\left( Hb^{\alpha}_{\beta}-H^2\delta^\alpha_\beta \right)$ and 
$\nabla^2\bn\cdot\nabla^\beta\bn = \nabla_\alpha\left( \nabla^\alpha\bn\cdot d_\beta\bn - \tfrac{1}{2}\nabla^\gamma\bn\cdot d_\gamma\bn \, \delta^\alpha_\beta \right)$. An analog of the latter, which is just a simple consequence of the ability to permute derivatives (mentioned in Section \ref{notation}), will be used below.

The tangential part of \eqref{divH} is the gradient of a scalar or, equivalently, the %surface?
 divergence of an isotropic tensor.  %\com{(on the surface?)}. 
It is tempting to ascribe physical meaning to this, given that the Helfrich energy is fluid.  This tensor also seems at first glance to arise only from the constraint. That the divergence of the bending contribution seems to not have a tangential component was noted in \cite{CapovillaGuven02} and \cite{Powers10} (see also the appendix of \cite{Powers02}, and \cite{GuvenMuller08}), with the latter invoking certain invariance properties of the energy as explanation, but this seems to be a conflation of reparameterization invariance and stationarity under tangential displacements, of the sort criticized in \cite{AgrawalSteigmann09}. 
%who criticize the physicists' conflation of 
%\comb{ Powers10 "there is no tangential component of the force per unit area because the bending energy is invariant under changes in coordi- nates, and a small change in coordinates corresponds to a deformation of the surface along tangent directions at every point" not really true, see AgrawalSteigmann09 ``while it is appropriate to impose the invariance of the energy under arbitrary, hence infinitesimal, reparametrizations of the surface, in general this is not equivalent to rendering it stationary under tangential displacements ualpha of material points.''. }.
Upon closer inspection, both of these features of the tangential term--- isotropy and absence of bending contribution---  
may be physically irrelevant accidents that occur with this particular geometric energy. 
 It is potentially confusing when geometric quantities describe both the energy, and the geometry, of objects. 
%\com{It is certain that} 
The common arrangement of the term into the form \eqref{divH}, aided both by the Codazzi relations and the ability to move terms around that are proportional to pieces of the energy density, belies a more fundamental structure.  
There is an important distinction between the tangential part of the stress ${\mathbf f}^\alpha$ and the tangential part of the divergence of the stress. The tangential part of the stress will balance forces applied on boundaries, such as cuts in, or edges of, the membrane. The divergence of the stress appears in the balance of momentum, whose tangential projection can be rearranged into the balance of pseudomomentum. In the current setting, this is a conservation law; the vanishing of the (referential) divergence of a ``material stress'' (also known as a \mbox{pseudo-,} Eshelby, or configurational stress, or the non-temporal part of an ``elastic energy-momentum tensor'' \cite{eshelby75, peierls1985, gurevichthellung90, nelson91, maugin1993, gurtin00, yavari06, SinghHanna21}), 
 is a consequence of the relabeling symmetry of a homogeneous material. %whether fluid or solid.
% \com{which appears to be what some physicists are calling reparameterization invariance}.  
External forces, whether body forces or forces applied from one or the other side in the ``interior'' of the membrane, appear as source terms in the momentum balance but can be absorbed into the divergence in the pseudomomentum balance, whose source terms only come from broken material symmetries such as inhomogeneities in properties \cite{herrmannalicia1981, maugin1993, yavari06, SinghHanna21}.  
%\com{so the discussion around equation A10 of Powers02 is potentially misleading and/or a misinterpretation of this...}
%\com{ I do not know what to make of CapovillaGuven02 analogy with Bianchi identities},
%\comb{CapovillaGuven02 say "One can think of these equations as the Bianchi identities associated with surface reparametrizations.”} 
 
 \newpage
For the sake of comparison and illustration, we rearrange the tangential projection of the divergence of the area-constrained Atluri/SN stress \eqref{fSNa}. 
From Appendix \ref{SNstress} we know that, given $J=1$, we can write ${\mathbf f}^\alpha_\text{\tiny (SNa)} = - \tfrac{1}{2}\nabla_\gamma\left( A^{\beta\gamma}b_\beta^\alpha \right)\bn + \sigma_\text{\tiny (SNa)}\ba^\alpha 
= \tfrac{1}{2}\bar\nabla_\gamma\left( A^{\beta\gamma}d_\beta\bn \right)\cdot\ba^\alpha\bn  + \sigma_\text{\tiny (SNa)}\ba^\alpha$, and therefore 
\begin{align}
	-\nabla_\alpha {\mathbf f}^\alpha_\text{\tiny (SNa)} \cdot\ba_\eta &= 
	-\tfrac{1}{2}\bar\nabla_\gamma\left( A^{\beta\gamma}d_\beta\bn \right) \cdot d_\eta\bn 
	- d_\eta\sigma_\text{\tiny (SNa)} \, , \nonumber \\
	&= \bar\nabla_\gamma\left( \tfrac{1}{2}A^{\beta\gamma}d_\beta\bn\cdot d_\eta\bn - \tfrac{1}{4}A^{\alpha\beta}d_\alpha\bn\cdot d_\beta\bn\,\delta^\gamma_\eta - \sigma_\text{\tiny (SNa)}\delta^\gamma_\eta \right) \, , \\
	&= \bar\nabla_\gamma\left( \tfrac{1}{2}A^{\beta\gamma}b_\beta^\alpha b_{\alpha\eta} - \tfrac{1}{4}A^{\alpha\beta}b_\alpha^\zeta b_{\beta\zeta} \,\delta^\gamma_\eta - \sigma_\text{\tiny (SNa)}\delta^\gamma_\eta \right) \, , \nonumber
\end{align}
using an identity analogous to one mentioned earlier for Helfrich.  
This more clearly reveals the Eshelbian (material energy-momentum) form, in which the energy density appears subtracted from another quantity. This structure is always present, but here, unlike in the previous example, we cannot use Codazzi to simplify or rewrite in terms of gradients of scalar quantities. 
For completeness, the normal projection obtained from this model is %shape equation
\begin{align}
	-\nabla_\alpha {\mathbf f}^\alpha_\text{\tiny (SNa)} \cdot\bn = \tfrac{1}{2}\nabla_\alpha\nabla_\gamma\left(A^{\beta\gamma}b_\beta^\alpha\right) - 2H\sigma_\text{\tiny (SNa)}\,.\label{divSNanorm}
\end{align}  
The rudimentary example in the following Section \ref{rudimentary} highlights a difference between  \eqref{divSNanorm} and the normal projection of \eqref{divH}, also making use of results from the previous Section \ref{energies}.

\section{Example: Unidirectional bending}\label{rudimentary}

A rudimentary example, using results from Sections \ref{energies} and \ref{eshelby}, serves to illustrate important differences between the fluid Helfrich \eqref{helfrich} and solid area-constrained Atluri/SN bending \eqref{SNa} models.  
 %${\mathbf f}^\alpha_\text{\tiny (H)}$ and area-constrained ${\mathbf f}^\alpha_\text{\tiny (SNa)}$ 
%what we are really doing is considering a stress tensor $\ba_\alpha {\mathbf f}^\alpha$ and obtaining its traction by dotting with ${\bm \nu}$
%boundary tractions $\nu_\alpha{\mathbf f}^\alpha_\text{\tiny (H)}$ and $\nu_\alpha{\mathbf f}^\alpha_\text{\tiny (SNa)}$
For simplicity of calculation, we consider a patch of membrane uniform in curvature and metric, such that terms involving derivatives of these vanish.  
The condition on curvature is actually rather restrictive, leading to discrete special cases of unbent planar, isotropically bent spherical, or unidirectionally bent cylindrical patches, with only the last being anisotropic in curvature. 
The cylindrical example includes the physical case of the tether mentioned in Section \ref{fluidity}, 
%discussed in Section \ref{discussion}, 
and displays the relative complexity of the Helfrich response and simplicity of the Atluri/SN response.  

The multipliers contributing to the isotropic part of the tractions can be specified by balancing the normal component of the divergence of stress from \eqref{divH} or \eqref{divSNanorm} against a suitably nondimensionalized pressure difference $\Delta P$ between the two sides of the membrane. 
%across the patch. % (we have not introduced a bending modulus in our discussion as there were no other forces...) 
For a uniform patch, we obtain $\sigma_\text{\tiny (H)} = H^2-K+\Delta P/2H$ and $\sigma_\text{\tiny (SNa)}=\Delta P/2H$. 

The tractions $\nu_\alpha{\mathbf f}^\alpha$ at each point on the boundary of a deformed patch 
can be considered in terms of their components on the local orthogonal unit triad consisting of the surface normal $\bn$ and two tangential directions, the tangent normal ${\bm \nu}$ $(= \nu_\alpha\ba^\alpha)$ and tangent tangent ${\bm \tau}$.  %$(=\tau_\alpha\ba^\alpha)$ 
Consider a uniform cylindrical patch, curved in a single direction.  From the stresses \eqref{fhelfrich} and \eqref{fSNa}, we see that the normal component $\nu_\alpha{\mathbf f}^\alpha \cdot \bn = 0$ for either model, and we further obtain 
$\nu_\alpha{\mathbf f}^\alpha_\text{\tiny (H)} \cdot {\bm \nu} = -Hb_{\nu\nu}+\sigma_\text{\tiny (H)} + H^2$, 
$\nu_\alpha{\mathbf f}^\alpha_\text{\tiny (H)} \cdot {\bm \tau} = -Hb_{\nu\tau}$, 
$\nu_\alpha{\mathbf f}^\alpha_\text{\tiny (SNa)} \cdot {\bm \nu} = \sigma_\text{\tiny (SNa)}$, and
$\nu_\alpha{\mathbf f}^\alpha_\text{\tiny (SNa)} \cdot {\bm \tau} = 0$ (with these local orthonormal $\nu$-$\tau$ coordinates, the index type is immaterial, and $2H=b_{\nu\nu}+b_{\tau\tau}$). 
This situation is shown in Figure \ref{patch}; not shown are pressures normal to the membrane on either side, and moments around the tangent tangent, also necessary to sustain equilibrium.
%on a uniform (curvature and stretch) patch (derivative terms vanish as does any traction normal to patch). 

\begin{figure}[h]
	\includegraphics[width=5in]{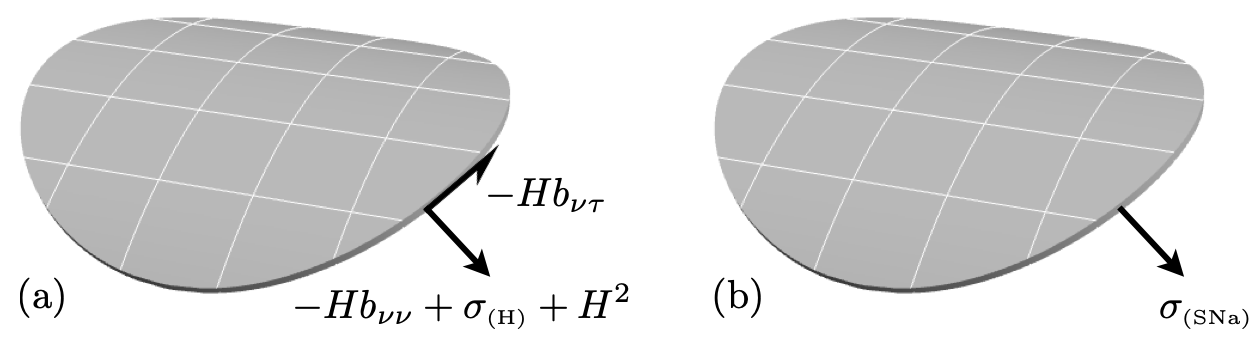}
	\captionsetup{margin=0.5in}
	\caption{Components of boundary tractions $\nu_\alpha{\mathbf f}^\alpha$ on a uniform patch of unidirectionally-curved membrane corresponding to two area-constrained bending energy models, (a) Helfrich and (b) Atluri/SN, shown at one location on the boundary and expressed in local orthonormal coordinates there. 
	Uniformity eliminates terms involving derivatives, and also components normal to the membrane, leaving only the in-plane tangent normal $\nu$ and tangent tangent $\tau$ (shear) components, with simplified forms. 
Not shown are pressures normal to the patch on either side, and moments around the tangent tangent, also necessary to sustain equilibrium. 
The fluid Helfrich tractions are anisotropic, while the solid Atluri/SN is tangentially isotropic. 
}\label{patch}
\end{figure}

The Atluri/SN solution is particularly simple.  There is a tangentially isotropic (pressure-like) stress.  Furthermore, if there is no difference $\Delta P$ in external normal pressures, then $\sigma_\text{\tiny (SNa)}$ must be zero (as $H$ is nonzero), and there are no tangential tractions at all.  The relation between moments and bending is decoupled from the in-plane stresses. 
The Helfrich solution is more complex.  The tractions feature anisotropic contributions, 
whose appearance in the tangent normal or tangent tangent (shear) directions depends on the local orientation of the boundary. 
%Traveling around the boundary of a uniform patch is akin to 
%The situation is akin to that of an infinitesimal patch representing a material element 
The curvatures $b_{\nu\nu}$ and $b_{\nu\tau}$ vary around the boundary.  
At four points corresponding to two principal directions, $b_{\nu\nu}$ takes on its extreme values and $b_{\nu\tau}$ vanishes.  A similar statement can be made about the in-plane stresses themselves. 
For such details of the behavior of symmetric two-dimensional tensors, we refer the interested physicist to the exposition of Mohr’s circle in an undergraduate engineering text such as \cite{UguralFenster}; however, to reproduce an analogously simple picture of a stressed small square of material for an arbitrary curved surface, the derivative terms in the tractions, including normal components, need to be considered. 

Returning to the case of the pulled membrane tether, we note that Powers and co-workers \cite{Powers02}, like others since,
 assumed that such experimentally realized tethers are cylindrical, with no pressure difference between the two sides of the membrane.  
They pointed out that the anisotropy of the Helfrich stress is necessary to construct this solution while satisfying a boundary condition corresponding to a pulling force.  
%expts to measure bending stiffness are consistent with the model and the assumptions, one sees the right force scaling with tension. Cuvelier05, Shi18
%helfrich admits cylindrical sols with no pressure difference and a pulling boundary condition (which is what people have been assuming tethers are) see also derenyi02, Fournier07, monnier10, Sahu20(scriven-love) 
For a cylindrical shape ($K=0$) and no pressure difference ($\Delta P=0$), we find $\sigma_\text{\tiny (H)} = H^2$ in the Helfrich solution.  
Inserting values, we observe that there is no traction on portions of the boundary whose tangent normals are in the bending direction, but on portions of the boundary in the orthogonal direction, a tangent normal component of $2H^2$ is required to sustain equilibrium. 
There is an applied force in the unbent (axial) direction, while the circumferential (hoop) stress is zero.  
On the other hand, the Atluri/SN bending energy does not admit such solutions, requiring either a pressure difference, or a non-cylindrical shape. % (see Appendix \ref{rudimentary} for analysis).   
For this bending energy, we find 
% for the Atluri/SN bending energy, %Mohrs circle of stress and curvature seem complementary here... 
 $\sigma_\text{\tiny (SNa)}=0$, which precludes the existence of any applied force; the cylindrical tether with an axial force and no pressure difference is not a solution.  More generally, the applied force and pressure difference are directly linked.

\section{Discussion and Conclusions}\label{discussion}

The existence or absence of shear stresses is a consequence of a constitutive choice, and the elegance and ease of definition of a geometric energy belies its complex response. 
Quadratic-curvature energies, such as Helfrich, are such that curvatures give rise to tangential stresses, and these are in general anisotropic. 
Energies quadratic in the product of stretch and curvature, such as Atluri/SN, simply do not have this coupling property.  
This result is only indirectly related to the requirement of fluidity, which restricts the quantities admissible in an energy to the Jacobian determinant and the geometric curvature invariants \cite{Steigmann99arma}. 
It may be possible to construct a fluid-like energy quadratic in the product of stretch and curvature that consists only of a term analogous to the curvature determinant; a term analogous to the square of the curvature trace seems to be precluded, although this remains an open question. 

While there is a stark qualitative distinction between these models, 
 the differences are nonetheless quantitatively only of the order of the in-plane strain.  This is small for stiff solids, for which there is a penalty for any type of strain, and with metric constraints the physical distinctions disappear altogether. 
However, constraining only the area does not limit the size of shear strains. In such cases, the differences in shear stress are of the order of the curvature, comparable to other stresses generated by bending. 

The perspective of pseudomomentum may be helpful in modeling heterogeneous vesicles, where the corresponding balance law will feature source terms.  %\com{CapovillaGuven04 say source terms show up in the shape equation, but I don't think this interpretation is right, not sure if we should comment} 
%instead of a conservation law. 

The present results regarding stresses and bend-stretch coupling are likely relevant to the modeling of elastic capsules in viscous flows \cite{Barthes-Biesel16}, which sometimes combines an area-constrained Helfrich energy with an additional penalty on surface stretching.  %(sometimes there is no bending though, as in BB work)
Within that literature, a variety of other models have been employed, including the complete neglect of bending resistance. Pozrikidis \cite{Pozrikidis01} posits a linear moment-curvature relation without coupling to in-plane stresses, assuming ``sufficiently small bending deformations''.  Without this restriction, the prescription is not consistent with variational derivation from a quadratic-curvature (or perhaps any) energy. He contrasts this with Zarda and co-workers'  \cite{zarda1977elastic} use of a Reissner-like axisymmetric shell bending measure to model red blood cells, and further recognizes that this measure is related to dilation invariance.  We note that a bending energy quadratic in the Reissner quantities would be akin to an Atluri/SN energy, and lead to a linear, decoupled constitutive response \cite{IrschikGerstmayr09, OshriDiamant17, WoodHanna19, vitral2023dilation, vitral2023energies}. % \comb{ ``The bending measures of strain (4.7) have been designed so that shape-preserving deformations do not induce bending moments; an example is the deformation of an expanding spherical shell.'' }

Our concern has been with elucidating the existence and origin of certain terms, and the structure of the equations.
%We remain agnostic on the value of any of these models in accurately representing any type of physical membrane, or their possible derivation from microscopic physics. 
Whether these models accurately represent any type of physical membrane, or are derivable from microscopic physics, are separate issues.  %A few relevant points follow. 
If the common assumptions surrounding tether formation are correct, so that shear stresses are required, another prevailing assumption is that these do not come from in-plane shear elasticity.  
%there is reason to think that these come from bending rather than in-plane shear elasticity.  
%Shear experiments on monolayers of relevant lipids at relevant temperatures show no evidence of solid-like elastic shear modulus \cite{Espinosa11}, and this is consistent with the observed translational and rotational freedom of membrane proteins \cite{Edidin74}. 
%indirect measurements/assumptions of a liquid phase on/in bilayers: EvansNeedham87
%\com{ The authors are not aware of similar direct measurements of shear of bilayers (in which they are not allowed to buckle out of plane)... nor of the bending energy of monolayers}
However, as summarized in \cite{Pamplona05}, Calladine and co-workers have suggested adding a small in-plane shear modulus, and coupling terms present in a ``thick'' shell with rest curvature, 
 %(which raises other complications related to inter-layer sliding) 
to resolve discrepancies between predictions derived using the Helfrich energy, with its bend-stretch coupling, and those derived using a direct model with linear, decoupled response, like that of Pozrikidis above.  The decoupled direct model fails to explain cylindrical tether pulling and oblate buckling phenomena.  
 %thick shell explanations of tethers: WaughHochmuth87, CalladineGreenwood02 
 %shear elasticity explanations of oblate/prolate buckling: PamplonaCalladine93 which originally suggested that a very small shear modulus should be present
%it seems there is an error of assumption here in Calladine's work, that the Helfrich model will give rise to a “first approximation” theory a la Koiter (see e.g. Acharya). 
The restrictions on the form of a fluid membrane energy are based on the presumption of in-plane fluidity, not three-dimensional fluidity of a thin bulk film of fluid \cite{Deseri08}.  A simple film of fluid would not have a bending energy, including aspects of a static bending response such as transverse shear stresses. % (the authors are not aware of a published dimensional reduction, which would anyway not be relevant to a lipid bilayer--- however see integrations such as those by Ribe02). 
The Helfrich model assumes that the membrane's material properties endow it with an elastic bending energy while still retaining in-plane fluidity. %, whether it be splay energy, coupling between leaflets, etc.
The biophysics literature provides some suggestions of microscopic origins for bending resistance of a bilayer \cite{JanmeyKinnunen06}. 
%``Membranes resist bending because changing local curvature alters both the headgroup spacing and the entropy of the hydrophobic chains.'' 
Red blood cells (a type of capsule) have a small shear modulus, associated with an additional spectrin-actin network \cite{HochmuthWaugh87, JanmeyKinnunen06}. 
%%indirect measurements using red blood cells: very small but nonzero shear modulus, see e.g. WaughEvans79, Evans73-2, Lelievre95, Henon99... however it is assumed this comes from spectrin network not from the bilayers. 
However, Waugh \cite{Waugh87} found that red blood cells with spectrin abnormalities also have a reduced bending stiffness correlating with their reduced shear stiffness, implying a significant solid contribution to their bending energy. 

The Helfrich energy has been successfully applied to explain static and dynamic behaviors of structures formed from closed lipid bilayer membranes. 
The only problems with its use that we warn against are the misinterpretation of shear stresses, and the ``validation'' of analytical results from Helfrich using SN-type numerics whose continuum limit is a different energy with different qualitative behavior. 
%Helfrich also does well in modeling various dynamics of vesicles although probably other bending energies would do so as well. 

%Shear stresses in membranes with bending elasticity were discussed in the context of a few simple and common fluid and solid models, alongside the structure of the momentum and pseudomomentum balances, and some commentary on the literature. 

\newpage

\appendix

\section{Calculation of the Atluri/SN stress}\label{SNstress}

The only piece of the calculation not already available in the literature is the derivation of the stress arising from the Atluri/SN bending energy that feeds into \eqref{fSNa} and \eqref{fSNm}. 

The below derivations employ the symmetry of metric and curvature tensors, certain properties of derivatives mentioned in Section \ref{notation}, the Weingarten relations, and the relations 
 $\delta\bn = -\bn\cdot\delta\ba_\alpha\ba^\alpha$ and $\delta b_{\alpha\beta} = \delta b_{\beta\alpha} =  \nabla_\alpha d_\beta\delta\bx\cdot\bn$.  %\com{cite or derive?}. 
 These results are not fully general, as they also employ the constraint $J=1$, which trivializes the Piola relations between divergences. 
Either 
\begin{align}
	\mathrm{with}\; J=1\,, &\;\;\,\delta \int \!\mathrm{d}A\,  \tfrac{1}{4}A^{\gamma\beta} b^\alpha_\gamma b_{\alpha\beta}  \nonumber \\
	&=\int \!\mathrm{d}A\, \tfrac{1}{4}A^{\gamma\beta} \delta \left( a^{\alpha\eta}b_{\gamma\eta}b_{\alpha\beta}  \right) \, , \nonumber \\
	&=\int \!\mathrm{d}A\, \tfrac{1}{2}A^{\beta\gamma} b^\alpha_\beta \left( d_\gamma\bn\cdot \delta\ba_\alpha + \nabla_\gamma \delta\ba_\alpha\cdot\bn \right) \, , \nonumber \\
	=\text{(boundary terms)} &- \int \!\mathrm{d}A\, \bar\nabla_\alpha \left[ \tfrac{1}{2}A^{\beta\gamma}b_\beta^\alpha d_\gamma\bn  - \tfrac{1}{2}\nabla_\gamma\left( A^{\beta\gamma}b_\beta^\alpha \bn \right) \right] \cdot\delta\bx \, , \nonumber \\
	 =\text{(boundary terms)} &- \int \!\mathrm{d}A\, \bar\nabla_\alpha \left[ - \tfrac{1}{2}\nabla_\gamma\left( A^{\beta\gamma}b_\beta^\alpha \right)\bn \right] \cdot\delta\bx \, , \nonumber
\end{align}
or alternately 
\begin{align}
	\mathrm{with}\; J=1\,, &\;\;\,\delta \int \!\mathrm{d}A\,  \tfrac{1}{4}A^{\gamma\beta} d_\gamma\bn\cdot d_\beta\bn  \nonumber \\
	&=\int \!\mathrm{d}A\, \tfrac{1}{2}A^{\beta\gamma} d_\beta\bn \cdot d_\gamma\delta\bn \, , \nonumber \\
	&=\int \!\mathrm{d}A\, \tfrac{1}{2}A^{\beta\gamma} b_\beta^\eta\ba_\eta \cdot d_\gamma(\ba^\alpha\bn\cdot \delta\ba_\alpha) \, , \nonumber \\
	 =\text{(boundary terms)} &- \int \!\mathrm{d}A\, \bar\nabla_\alpha \left[ - \tfrac{1}{2}\bar\nabla_\gamma\left( A^{\beta\gamma}b_\beta^\eta\ba_\eta \right)\cdot\ba^\alpha\bn \right] \cdot\delta\bx \, , \nonumber \\
	 =\text{(boundary terms)} &- \int \!\mathrm{d}A\, \bar\nabla_\alpha \left[ - \tfrac{1}{2}\nabla_\gamma\left( A^{\beta\gamma}b_\beta^\alpha \right)\bn \right] \cdot\delta\bx \, . \nonumber
\end{align}
We note that a cursory comparison between the expression $\nabla_\gamma\left(A^{\beta\gamma}b_\beta^\alpha\right)$ appearing here, which employs a present covariant derivative, and the expression $\bar\nabla_\gamma \mu^{\gamma\alpha}$ appearing in the stress in \cite[Equation 60]{vitral2023energies}, which employs a referential covariant derivative, suggests a disagreement.  However, it can be shown that the tangential projection of the angular momentum balance \cite[Equation 64]{vitral2023energies} requires that $\mu^{\alpha\beta}\left( \bar\Gamma^\gamma_{\beta\alpha} - \Gamma^\gamma_{\beta\alpha} \right)$ must vanish, and this can be used to rewrite the term 
such that $\bar\nabla_\beta\mu^{\alpha\beta}$ can be replaced with $J\nabla_\beta\left(J^{-1}\mu^{\alpha\beta}\right)$.  Note that this is not a consequence of a Piola identity; there is no such relation in component form for a divergence when there is a dangling index.

\newpage

\section*{Acknowledgments}
We thank TJ Healey, DJ Steigmann, and particularly E Vitral, for discussions prior to the initial draft, and M. Deserno and T. Powers for discussions afterwards. 
JAH was supported by U.S. National Science Foundation grant CMMI-2001262.

\bibliographystyle{unsrt}
%\bibliography{refs_hsn}

\end{document}